\documentclass[aps,prl,showpacs,showkeys,twocolumn]{revtex4-1}
\usepackage{graphicx,hyperref,amsmath,amsfonts}
\usepackage{epstopdf,color,bm,multirow,rotating,soul}

\begin{document}

\title{Towards broadband AFC photon echo quantum memory}

 \author{S.A.Moiseev and N.M.Arslanov }
 \affiliation{Kazan Quantum Center, Kazan National Research Technical University n.a. A.N.Tupolev-KAI, 10 K. Marx, Kazan 420111, Russia}
\date{\today}


\begin{abstract}
We investigated the optimal spectral design of periodic structure of narrow lines (AFC-structure) within the inhomogeneously  broadened atomic transition which are created  for the implementation of broadband AFC photon echo quantum memory.
The influence of the spectral design on the suppression of negative dispersion effects in the AFC-echo retrieval was studied for different spectroscopic parameters of atomic media.
The maps of the assigned spectral quantum efficiency have been constructed  for the created AFC-structures characterized by different spectral design, finesse and optical depth.
Based on the performed analysis, we discuss the possible ways for experimental implementation of highly efficient broadband AFC-protocol. 

\end{abstract}

\maketitle

\section{1.Introduction }

The spin/photon echo techniques  \cite{Hahn1950, Kopvilem1963, Kurnit1964} has demonstrated itself as a useful and widely used experimental methods of modern coherent laser spectroscopy. 
Moreover a number of interesting attempts have been also applied in elaboration of the photon echo based methods for storage and fast processing of optical information \cite{Zakharov1995, Kroll1993}, including  the  dynamical echo-holography \cite{Kroll1993, Shtyrkov1978}  and  nonstationary generalization of four wave mixing \cite{Shtyrkov1981, Mohan2000}. 
Recently it was shown that the photon echo can be applied for optical quantum memory (QM) \cite{Moiseev2001, Afzelius2010c}  which demonstrated high promise for the quantum storage of multi-qubit light fields in many successful experiments.
The multi-qubit QM is required by the creation of the universal quantum computer and quantum repeater of long-distance quantum communication.
Based on the proposal \cite{Moiseev2001,  Moiseev2003} called by CRIB, a number of different schemes are currently elaborated \cite{Kraus2006},(GEM) \cite{Alexander2006, Hetet2008c, Moiseev2008b}, (AFC) \cite{DeRiedmatten2008, Afzelius2009}, (ROSE) \cite{Damon2011},  (HYPER) \cite{McAuslan2011}, impedance matched QM \cite{Afzelius2010d, Moiseev2010d}, Raman-echo scheme \cite{Hosseini2011b, Moiseev2013}  multi-resonator schemes \cite{Moiseev2017, Moiseev2017a}. 

\begin{figure*}[htb]
\begin{center}
\includegraphics[width=0.95\textwidth]{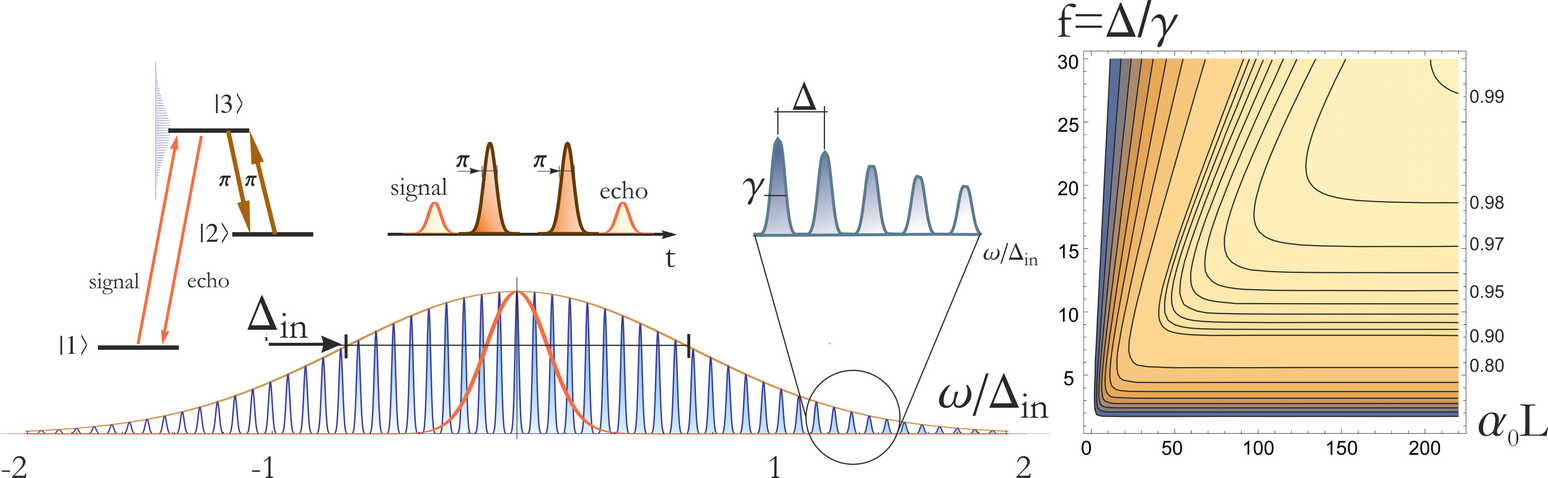}
\caption{\textbf{a)} AFC-structure of inhomogeneously  broadened line, depending on the optical density of the resonant transition; temporal scenario of AFC-protocol where two auxiliary laser pulses (rectangular) transfer an excited optical coherence on the long-lived levels and then return it back (first/last pulse is the signal/echo pulse); higher insert shows spectral shape of IB where $\Delta$-spectral distance, f-finesse, $\gamma$- linewidth of each comb, down insert shows level diagram of atoms and quantum transitions  for the signal/echo fields and for two additional  auxiliary laser pulses. \textbf{b)}~Map of quantum efficiency $\eta$ at  zero frequency in the line centre. Black lines denote the efficiency isolines for the efficiency values $\eta=10\%-99\%$.
}
\label{Fig1}
\end{center}
\end{figure*}

Below we focus to the AFC-protocol which has demonstrated practical advantages in operation with broadband light fields \cite{Usmani2010, Afzelius2009}. 
AFC approach provides an automatic  rephasing of the excited optical coherence and that significantly simplies an experimental realization of the photon echo approach and it pays the way for using short light pulses \cite{Afzelius2009} and for multiplexing of the light field storage \cite{Sinclair2014}.
Below we develop the approach to the high-efficiency broadband modified AFC-protocol and present optimal parameters for realistic experimental implementation.

\section{2. AFC-protocol}

AFC-protocol is implemented experimentally in systems of resonant atoms characterized by the inhomogeneous broadening (IB) of the optical resonant transition in a form of the periodic structure of narrow lines (Fig. \ref{Fig1}) . 
Usually, AFC-structure is created artificially in IB of resonant transition by using the  laser hole-burning technique. 
AFC/SFC-structure can be also realized in some atomic systems with  natural periodic structure of hyperfine splitting of the atomic transitions \cite{Gerasimov2014}.

\begin{figure}
 \centerline{\includegraphics[scale=0.2]{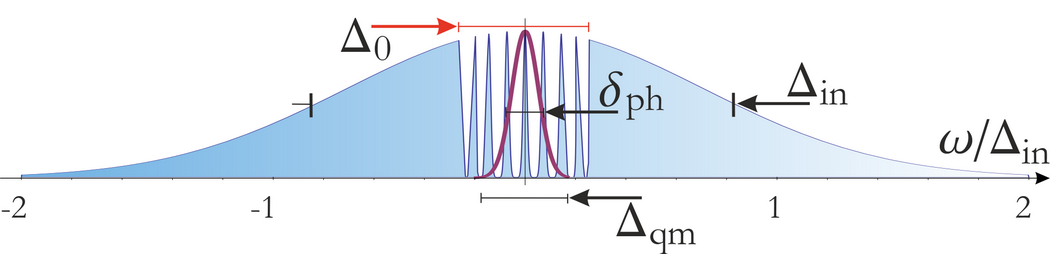}}
 \vspace*{8pt}
 \caption{ 
 The spectral quantum efficiency $\eta$ within the AFC-structure $\Delta_{qm}=0.1 \Delta_{in}$ prepared in a narrow inhomogeneously broadened line depending on the optical density of the resonant transition of $\alpha L$. A) efficiency $\eta$ is described by $\eta=|\Gamma(\omega)|^2$, $\Gamma(\omega)=\frac{1-exp{[i\omega_0\chi (\omega) L/c]}}{1-i\chi'(\omega)/\chi"(\omega)}$ in accordance with the expression (\ref{eta_AFC2}); b) For comparison, in this case, the efficiency $\eta=(1-e^{-\alpha L})^2$ does not take into account the influence of the dispersion effects (see(1)).
  }
 \label{Fig2}
 \end{figure}

AFC-protocol has been demonstrated in a number of inorganic crystals and classes doped by the rare-earth ions. 
The highest quantum efficiencies  have been  obtained in work \cite{Amari2010}, while practical usage of the quantum memory devices requires the quantum efficiency 90 \%  that stimulates further elaboration of this QM protocol. 
In order to achieve  these parameters in realistic systems characterized significant influence of transverse relaxation on the atomic transition it is reasonable to use sufficiently short light pulses. 
Herein, a spectrum of the short light pulses is limited by the inhomogeneous broadening  of the atomic transition $\delta\omega_f\sim\Delta_{in}$.
On the  Fig. (\ref{Fig1}) the map of the quantum efficiency $\eta$ for the center frequency of the optical transition  versus the optical density $a_0 L$ and the finesse $f$ \cite{Arslanov2017}. 
This case is suitable when the spectral width of the conserved pulse is small compared to the width of the AFC-structure.
As seen from Fig. (\ref{Fig1}b),in the existing 
experimental studies the crystals were used that allow the efficiency not exceeding 70\% \cite{Arslanov2017}. 
On the other hand  in the experiment, the burning of AFC-structure takes place in the certainly small region of the inhomogeneous broadening (Fig. (\ref{Fig2})). 
Therefore, the spectral sidepieces of resonant line could significantly influence to the efficiency  of the echo signal recovery.

However, the quantum efficiency  and fidelity of AFC-protocol decrease when spectral width of the signal pulse becomes comparable with inhomogeneous broadening  \cite{Moiseev2012} due to the negative influence of the dispersion effects.
Below we elaborate the highly efficient modified AFC-protocol where the negative dispersion effects can be strongly suppressed in a realistic way within wide spectral range so the high spectral quantum efficiency can be achieved for the quantum storage of short light pulses.  
Basic AFC-protocol \cite{DeRiedmatten2008, Afzelius2009}, uses atomic systems where IB is a periodic structure of narrow lines $\delta_j = n_j \Delta $ (where $ \delta_j \in \Gamma_{in} $, $ n_j = 0, \pm 1, \pm 2 ... $ - integers \cite{DeRiedmatten2008, Afzelius2009}, $ \Delta $ - the distance between the nearest two lines (see Fig. ~(\ref{Fig1}). 
After experimental preparation of such AFC-structure at time $ t = 0 $,  the atomic medium is excited by the signal light pulses.
Resonant absorption of the signal pulses is accompanied  by an excitation of the macroscopic atomic polarization $ P (t) $, which is disappeared  and is rephased automatically after the time  delay $ \tau = 2 \pi / \Delta $ leading to the echo-pulse irradiation \cite{DeRiedmatten2008, Chebotayev1983}.
It is worth noting that it is possible to interrupt  the rephasing of atomic coherence by reversible transfer of the atomic coherence on the long-lived spin coherence.
The described scenario  can be implemented for very short light pulses that make it very convenient for practice
in  comparison with other photon echo protocols which require an additional control technique.
To date, experimental works on the AFC-protocol has been implemented in the forward schemes of the photon echo retrieval that limits the quantum efficiency by the maximum theoretical limit  54\%.
The best quantum efficiency achieved in these studies was 35\% \cite{Amari2010, Zhou2013, Cho2016} 
(i.e. 65 \% of the maximum possible value). 
The placement of atoms in an optical cavity has allowed achieving greater quantum efficiency - 58\%.
However, using of the cavity with high-Q factor makes it impossible to store broadband light signals, as it can take place on the atoms in free space geometry.

Experimentally implemented simplified forward-scheme  convincingly demonstrated a great potential of AFC-protocol to store the broadband quantum states of light, however, further enhancement of quantum efficiency  requires a transition to the backward AFC-scheme.
In the scheme, the retrieved echo pulse propagates in the backward  direction, that promises achieving quantum efficiency close to 100 \cite {Afzelius2009} in accordance with 
$\eta\cong(1-e^{-\alpha_o L/f})^2e^{-7/f^2}$,
where $\gamma$ - linewidth of the combs, $\alpha_o L$ - the original optical depth of the resonant transition  before creation of AFC-structure, $L$- length of the medium, $ f=\Delta/\gamma$ - finesse, $\Delta$ is the spectral distance, in particular  $\eta \to 0.9$ for $\alpha_o L=40$ and $f=10$ \cite{Afzelius2009}. 

According to \cite{Afzelius2009}, the quantum efficiency $\eta$ depends on the optical depth and finesse of AFC-structure that is in an agreement with the experimental data \cite{Amari2010, Zhou2013, Cho2016}  obtained for forward AFC-protocol in atomic systems with relatively low optical depth and with low quantum efficiency, respectively.
There is only the optimal value of $f$ to get the maximum quantum efficiency for the given optical depth $\alpha_0 L$ of the optical transition.
However the optimal finesse does not give any information  about spectral properties of quantum storage.
It was recently observed \cite{Moiseev2012} that the efficiency of the backward AFC-protocol is described well $\eta$ (without dispersion), unless the signal field has a spectral width negligible narrow in comparison with IB linewidth $\Delta_{in}$ of the AFC-structure  Fig.(\ref{Fig1}).
However, in the case where the signal field has the spectral width of AFC-structure which is comparable to the IB  $\Delta_{in} $, the efficiency of AFC-protocol decreases due to the negative impact of the dispersion effects.
The dispersion effect in AFC-protocol has the fundamental causes determined by the lack of strict time reversibility in the emission of light-echo signal with respect to the absorption of  the input signal light pulse. 
It worth noting that in contrast to this case, for example, the original photon echo protocol (CRIB)  \cite{Moiseev2001} is time reversible  for an arbitrary spectral width of the signal light pulses \cite{Kraus2006, Moiseev2004b} that excludes the negative dispersion for arbitrary spectral width of the signal pulses. 
The lack of an accurate reversibility leads to violation of the phasematching for the emission of a photon echo signal, which leads to a decrease in quantum efficiency of AFC-protocol.
The mis-phasematching is realized  differently for  all  spectral components of the photon echo pulse, so the perfect phasematching cannot be restored completely  by the choice of the wave vectors of the additional control laser pulses which cause a radiation of the echo signal in the opposite direction.
This situation poses a serious experimental problem of increasing the spectral width of the signal in the AFC-protocol \cite{DeRiedmatten2008}.
The paper \cite {Moiseev2012} provides a method for solving this problem.
Below we show how it can be effectively implemented using the natural shape of inhomogeneously broadened lines of the optical transition.

\section{3.Broadband modified AFC-protocol}

\subsection{Backward scheme. Basic picture}

The backward AFC-protocol is realized in three steps which is similar to the original time-reverse scheme \cite{Moiseev2001}. 
At the beginning the signal pulse is absorbed by the atoms with an excitation of optical coherence $e^{-i(\omega t- \vec{k}_p \vec{r})} $ characterized by a set of different  wave vectors $\vec{k}_p = (\frac{\omega} {c} + \delta k) \vec{e}_z $ in IB atomic transition, where $\delta k = \frac{\omega_0 \chi (\delta)}{2c} $ determines an additional term caused by the resonant interaction of light with  atoms, $\omega_0 $ - absorption line center, $\Delta = \omega-\omega_0 $ - frequency detuning.
$\chi = \chi `+ i \chi``$ - susceptibility, where $\omega_0 \chi``/c = \alpha$ specifies the coefficient of resonant absorption, and $\chi '(\Delta)$ is an anomalous spectral dispersion.
The anomalous dispersion can lead to the  superluminal or negative group velocity $ \upsilon $ in the propagation of the light in a high optical density media $\alpha L \gg1 $   \cite{Wang2000}.
The spectral dispersion plays a significant role in  phase-matching  between the light field modes and atomic excitations in condensed media. 
After the signal light absorption, the excited atomic coherence (polarization) is transferred to the long-lived electron spin coherence by the action of controlled first (writing) laser pulse. 
The created spin wave stores information about anomalous spectral dispersion of the signal light pulse propagation in spin  wave vector $\vec{k}_p-\vec{k}_1 $ (where $\vec{k}_1$ - wave vector of the laser field).
It is important that the lifetime of the spin quantum coherence can be increased up to several hours in rare earth ions system. 

In the retrieval (third) step, the reading laser pulse with the wave vector $ \vec {k} _2 $ restores optical wave polarization of atoms $ e ^ {- i (\omega t- \vec{k}_{ret} \vec{r})} $ with new wave vector $\vec{k}_{ret} = \vec{k}_p-\vec{k}_1 +\vec{k}_2$.
Rephasing  of the macroscopic optical coherence (polarization) is similar to the forward AFC-protocol, but with an opposite wave vector of the atomic polarization leading to the emission of the echo signal in the backward direction if the wave vector of laser pulse $\vec{k}_1 =\frac{\omega}{c}\vec{e}_z$, $\vec{k}_2 = - \frac{\omega}{c}\vec{e}_z $. 
Effective recovery of the recorded information requires a sufficiently perfect fulfillment  of phase matching for all spectral components of light waves and waves of AFC-polarization.
The reradiated backward AFC echo field $A_{echo}(t, z)$ is described by the following expression \cite {Moiseev2012}:

\begin{align}
\label{eta_AFC2}
A_{echo}(t,z)=k(2\pi/\Delta) \int_{-\infty}^{\infty}\frac{d\omega}{2\pi}\Gamma(\omega)e^{-i\omega(t-z/c)}A_s(\omega),
\end{align}

\noindent
where $A_s(\omega)$ is spectral component of the signal field  at the entrance to the crystal, 
$\Gamma(\omega)=\frac{1-exp{[i\omega_0\chi (\omega) L/c]}}{1-i\chi'(\omega)/\chi''(\omega)}$ - spectral response function  (SRF) of backward AFC-protocol,
$\chi=\chi'+i\chi''$, 
$ \chi' $ - refractive index of the crystal structure of c AFC,
the absorption coefficient of the crystal  $\alpha=2\frac{\omega_0}{c}\chi’’=\frac{d}{LF}$,
$\kappa (2 \pi / \Delta) $ is the factor determining the existence of irreversible polarization decay in a time of $ 2 \pi / \Delta $, which is caused by the finite spectral width AFC combs (with a Gaussian line shape comb $ k = e^ {- 3.5 / F^2} $, which is implicit in Eq.(\ref{eta_AFC2}) ).

As can be seen from Eq.(\ref{eta_AFC2}), SRF predeterminates the frequency dependence of the quantum efficiency $\eta (\omega) = | \Gamma (\omega) | ^ 2 $ for sufficiently large finesse $F$ and weak decoherence $k\cong 1$, respectively.
Note that the spectral absorption coefficient of the signal field is determined by the parameters of the resonance absorption lines in this spectral region.
However, the spectral  behavior of dispersion $\chi '(\omega)$ also depends on the resonant lines, which exist in the immediate vicinity to this frequency domain. 
This is an important property of AFC-protocol, which is used in our approach for considerable improvement of spectral quantum efficiency in this protocol.  
It is noteworthy that, according to Eq. (\ref{eta_AFC2}), the impact of spectral dispersion is negligible only for sufficiently narrow spectral range $\chi'(\omega)/\chi''(\omega)\ll 1$ and for all the spectral range of AFC-structure but low optical depth of the atomic transition ($\alpha(\omega)L<1$), wherein  $\eta(\omega)\cong\frac{1}{4}{(\alpha(\omega)L)}^2\ll 1$ and low quantum efficiency, respectively.
The property of Eq.(\ref{eta_AFC2}) explains properly the experimental results \cite {Amari2010, Zhou2013, Cho2016}.

For significant broadening of the QM spectral range, we consider the case of AFC-structure created within spectral interval $\Delta_{0}<\Delta_{in}$ of IB with typical Gaussian spectral shape.
Below we study the optimal choice of $\Delta_{o}$ providing a significantly increase the spectral interval  $\Delta_{qm}$ of highly efficient QM.
In accordance with $\eta$ (without dispersion), quantum efficiency $\eta(\omega)>0.9$ is possible for the atomic systems with sufficiently high optical depth.
However, in this case, the negative effects of spectral dispersion on the quantum efficiency are  enhanced.

\subsection{Numerical simulation of  broadband modified AFC-structure}

\begin{figure*}[htb]
\begin{center}
\includegraphics[width=0.95\textwidth]{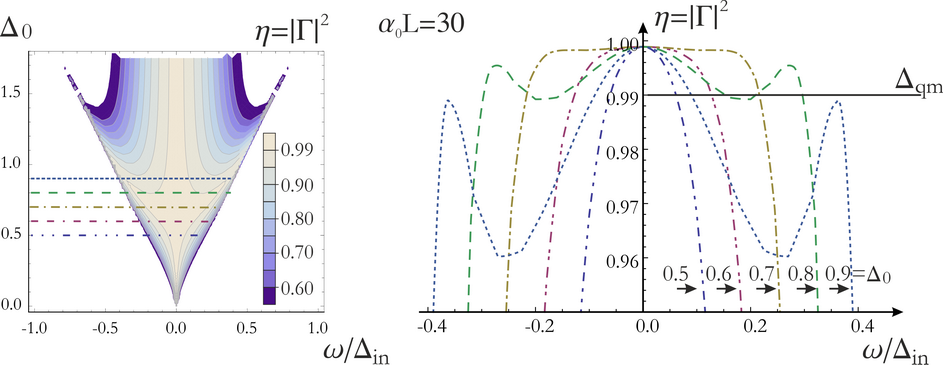}
\caption{ 
\textbf{a)} the map of spectral efficiency $\eta(\omega)$ vs. the width of the spectral interval $\Delta_0$  (in the units of $\Delta_{in}$) for the optical density $a_0L =30$ and $\omega$ within   $\Delta_0$.
\textbf{b)} the spectral efficiency $\eta (\omega)$ for the optical density $a_0L =30$ and $\Delta_{qm}  = 0.6 \Delta_{in}$  vs. $\omega / \Delta_{in} $  (the sections marked by line in Fig. (\ref{Fig3}a)); the arrow
shows the quantum efficiency behaviour at $\Delta_0 = 0.6 \Delta_{in}$.
 }
\label{Fig3}
\end{center}
\end{figure*}

Fig. (\ref{Fig3} a) shows the calculation  of spectral efficiency $\eta (\omega)$ for different spectral widths $\Delta_0$ of AFC-structure   (in units $\Delta_{in}$ ) for the optical depth $\alpha_o L =30$.
Such optical depth necessary for implementation  of highly efficient QM as it will be seen further. 
Fig. (\ref{Fig3}a) demonstrates an interesting dependence of high quantum efficiency  $\Delta_{qm}$ on the spectral width $\Delta_0$.
At some values of $\Delta_0$, there is a strong broadening of $\Delta_{qm}$  where $\eta (\omega)> 0.99$ ( $\Delta_0\sim 0.8$) so the area of high quantum efficiency has a shape of a pointed anchor in the 2D map of $\Delta_0$ and $\omega/\Delta_{in}$.
Thus the choice of optimal detuning $\Delta_0$ can strongly increase the amount of $\Delta_{qm}$.
The effect is a result of broadband suppression of the spectral dispersion effects caused by the opposite influence of the AFC-structure and two sidelines.
At small value $\Delta_0> 0.5$, the negative dispersion effects are caused by the sidelines leading to the steep spectral dispersion and slow light effect in AFC-structure.
While too large $\Delta_0> 1$ leads to the abnormal spectral dispersion and superluminal light field, respectively, also leading to the negative contribution of dispersion in $\eta (\omega)$.

The cross-sections with different $\Delta_0$ in Fig.(\ref{Fig3}a) are presented on an enlarged scale in  Fig.(\ref{Fig3}b) in order to characterize  the properties of high spectral efficiency. 
As it is seen in the figure, it is possible to get $\eta (\omega)>0.99$ within spectral range $\Delta_{qm}=0.6$ by creating AFC-structure $\Delta_0=0.8$.
It is also possible to have $\eta (\omega)>0.998$ for  AFC-structure within $\Delta_0=0.7$ but only within  narrower spectral range $\Delta_{qm}=0.36$.
Increase in detuning $\Delta_0>0.8$ worsens the spectral characteristics of quantum efficiency.  
Similar behavior occurs for different optical depths and higher optical depth increases minimal value $\Delta_{qm}$.
So, if we set  some minimum quantum efficiency $\eta(\omega)$ its imposes the corresponding requirement  to the optical depth $\alpha_o L =30$ and $\Delta_{0}$. 
These parameters determine a minimal value $\Delta_{qm}$, respectively.
For the robust experimental realization of highly efficient broadband quantum memory on AFC-structure it is important to know parameters of systems  $\Delta_0$ and  $\Delta_{qm}$ for the given values of  $\alpha_0 L$ and $\Delta_{in}$.
Below we construct a series  of maps for the spectral detunings $\Delta_0$, $\Delta_{qm}$, finesse $f$ and optical depth $\alpha_0 L$ which can be used for finding the parameters of the AFC-structure that will allow the optimal way to implement experimentally broadband QM with the quantum efficiencies: $\eta= 20\%, 30\%, 40\%, 50\%, 60\%,  70\%, 80\%, 90\%$.
In particular, we are interested in the conditions for implementation of  quantum storage in broadband spectral range $\Delta_{qm}\geq1$ and high efficiency up to $\eta=90\%$.

\begin{figure}
\centerline{\includegraphics[scale=0.3]{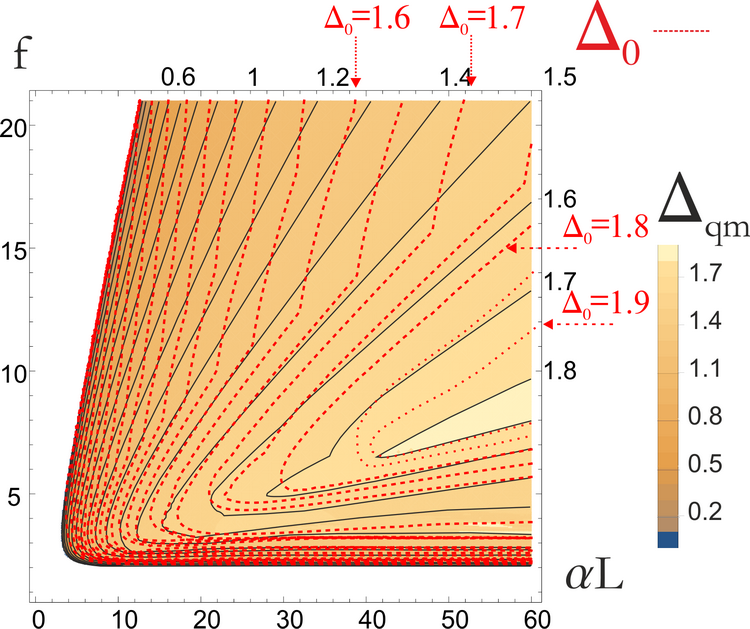}}
\vspace*{8pt}
\caption{ 
Map maximum width of $\Delta_{qm}$ with optimal $\Delta_0$ in units $\Gamma_{in}$. Red lines show contours of $\Delta_0^{opt}$ efficiency $\eta=20\%$.
}
\label{Fig4}
\end{figure}

 \begin{figure}
\centerline{\includegraphics[scale=0.3]{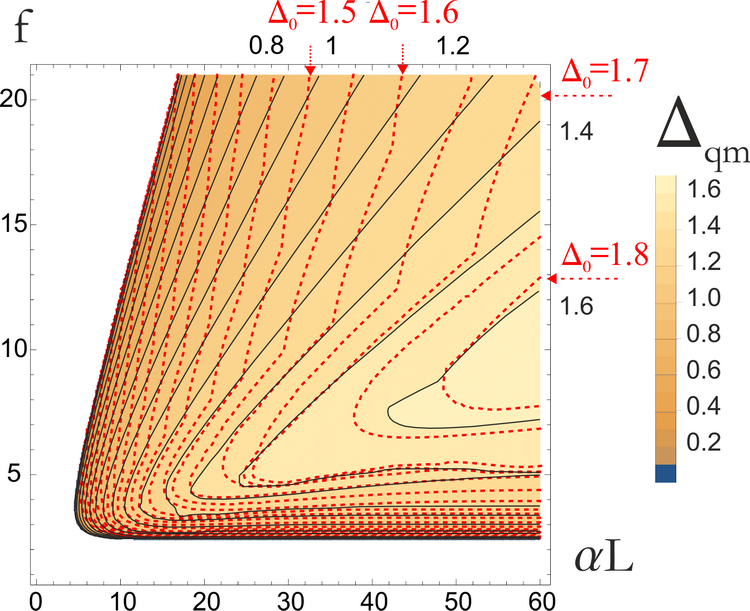}}
\vspace*{8pt}
\caption{ 
Map maximum width of $\Delta_{qm}$ with optimal $\Delta_0$ in units $\Gamma_{in}$. Red lines show contours of $\Delta_0^{opt}$ efficiency $\eta=30\%$.
}
\label{Fig5}
\end{figure}

 \begin{figure}
\centerline{\includegraphics[scale=0.3]{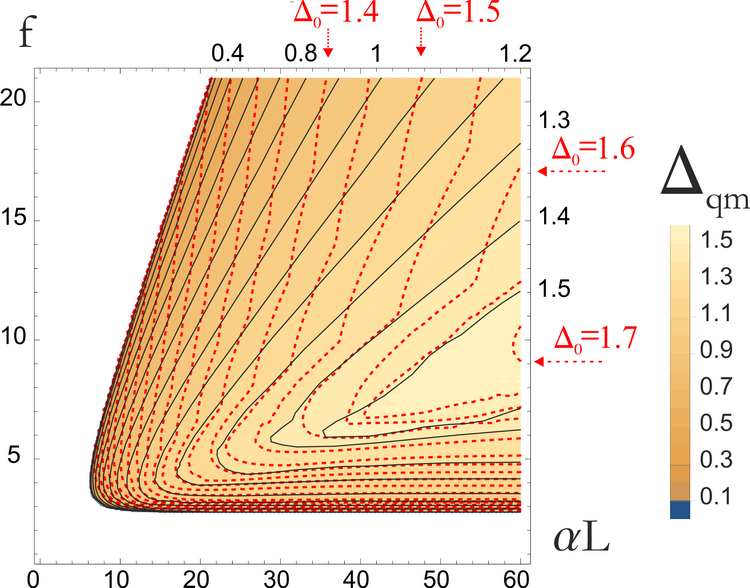}}
\vspace*{8pt}
\caption{ 
Map maximum width of $\Delta_{qm}$ with optimal $\Delta_0$ in units $\Gamma_{in}$. Red lines show contours of $\Delta_0^{opt}$ efficiency $\eta=40\%$.
}
\label{Fig6}
\end{figure}

 \begin{figure}
\centerline{\includegraphics[scale=0.3]{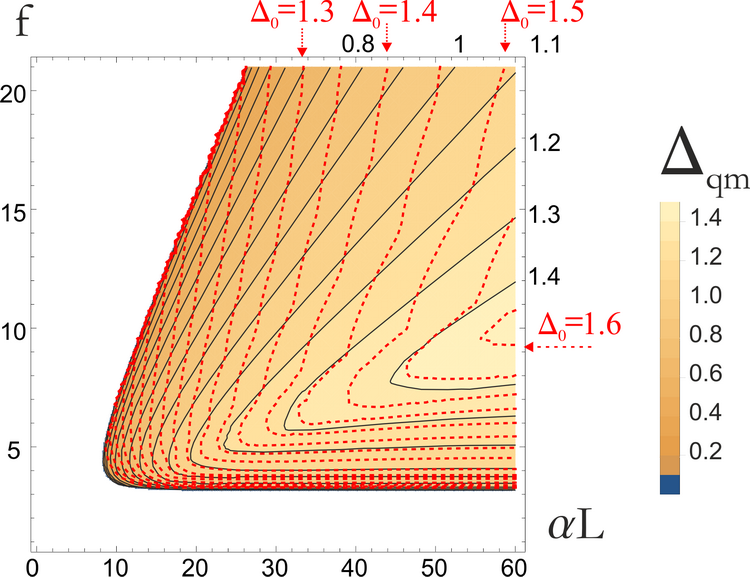}}
\vspace*{8pt}
\caption{ 
Map maximum width of $\Delta_{qm}$ with optimal $\Delta_0$ in units $\Gamma_{in}$. Red lines show contours of $\Delta_0^{opt}$ efficiency $\eta=50\%$.
}
\label{Fig7}
\end{figure}

 \begin{figure}
\centerline{\includegraphics[scale=0.3]{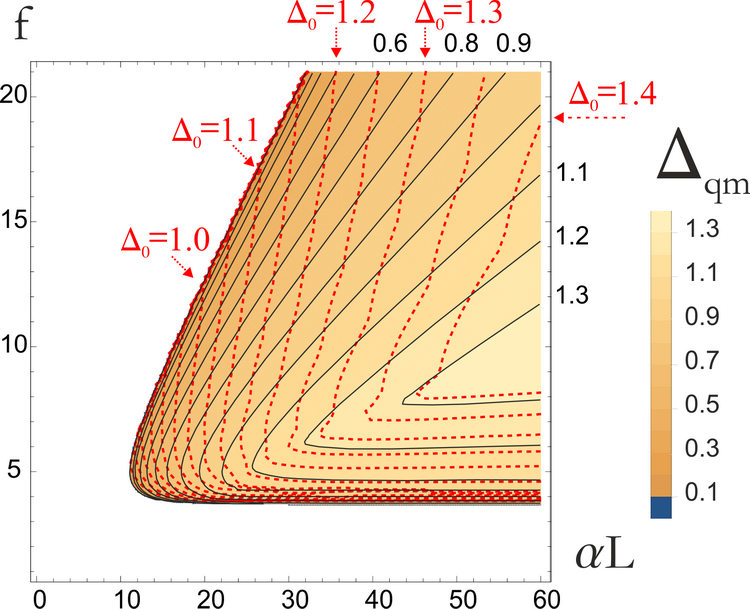}}
\vspace*{8pt}
\caption{ 
Map maximum width of $\Delta_{qm}$ with optimal $\Delta_0$ in units $\Gamma_{in}$. Red lines show contours of $\Delta_0^{opt}$ efficiency $\eta=60\%$.
}
\label{Fig8}
\end{figure}

\begin{figure}
\centerline{\includegraphics[scale=0.3]{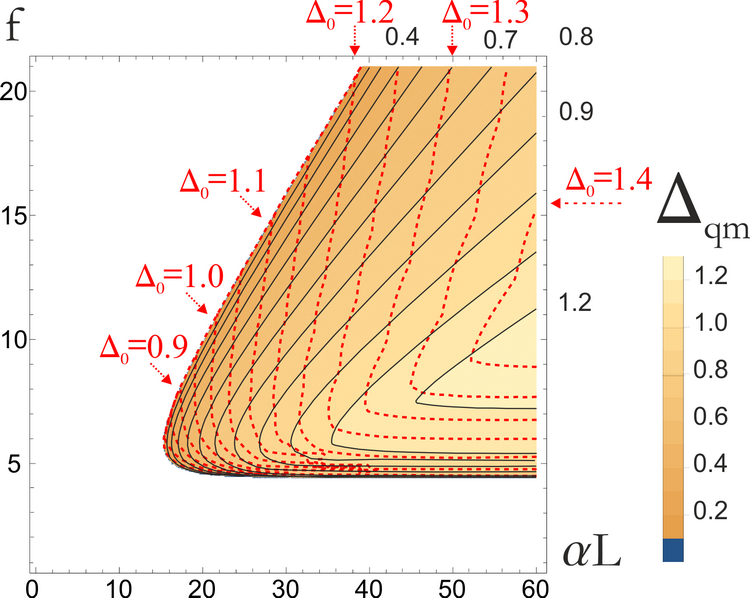}}
\vspace*{8pt}
\caption{ 
Map maximum width of $\Delta_{qm}$ with optimal $\Delta_0$ in units $\Gamma_{in}$. Red lines show contours of $\Delta_0^{opt}$ efficiency $\eta=70\%$.
}
\label{Fig9}
\end{figure}

\begin{figure}
\centerline{\includegraphics[scale=0.3]{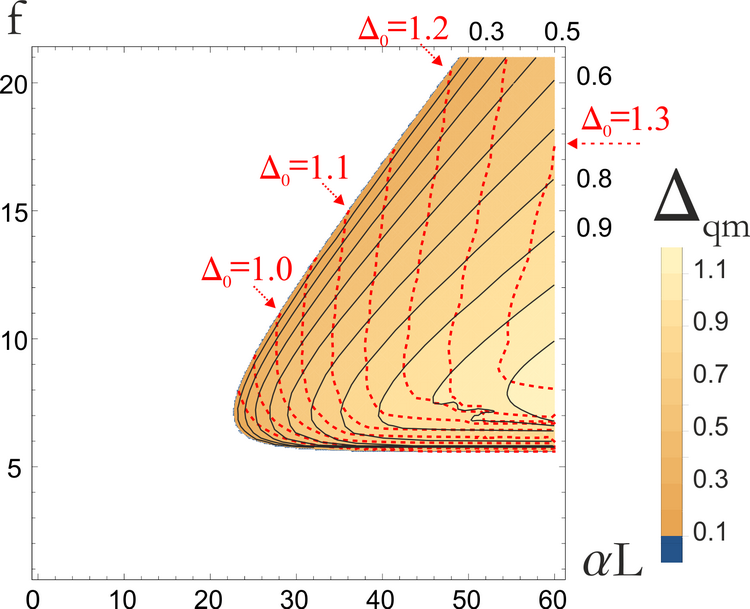}}
\vspace*{8pt}
\caption{ 
Map maximum width of $\Delta_{qm}$ with optimal $\Delta_0$ in units $\Gamma_{in}$. Red lines show contours of $\Delta_0^{opt}$ efficiency $\eta=80\%$.
}
\label{Fig10}
\end{figure}

\begin{figure}
\centerline{\includegraphics[scale=0.3]{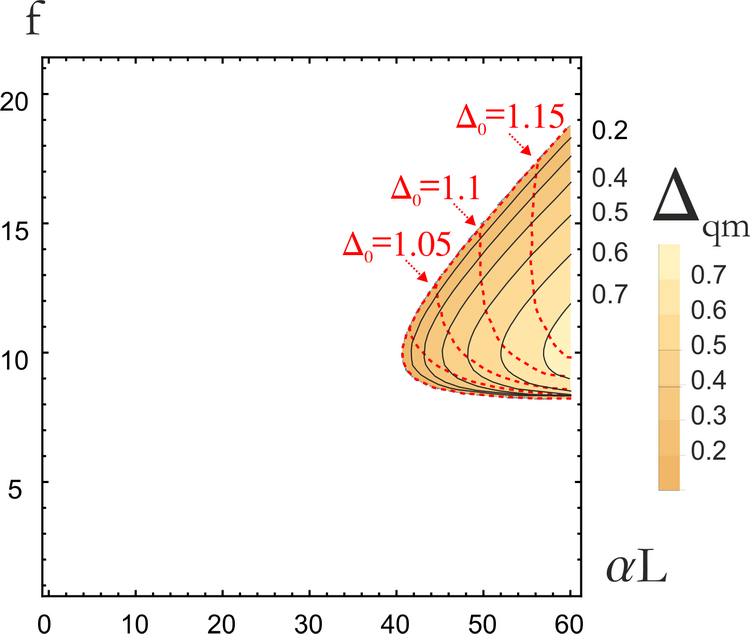}}
\vspace*{8pt}
\caption{ 
Map maximum width of $\Delta_{qm}$ with optimal $\Delta_0$ in units $\Gamma_{in}$. Red lines show contours of $\Delta_0^{opt}$ efficiency $\eta=90\%$.
}
\label{Fig11}
\end{figure}

\textbf{1. Map ($\eta= 20\%$)} Fig. (\ref{Fig4}).

For the given efficiency, the map shows possible values of optical depth $\alpha_0 L$, finesse $f$  and spectral detunings $\Delta_0$ and $\Delta_{qm}$ in units of $\Delta_{in}$. 
Specific values of $\Delta_0$ are depicted by the red dashed lines with $\Delta_0$ varied from $0.1$ up to $1.9$.
The realized spectral widths $\Delta_{qm}$ are shown by the solid black lines where $\Delta_{qm}$ is varied up to $1.8$. 
It is worth  noting that increase of finesse $f$ after some value $f_o$ worsens the spectral properties of quantum efficiency. 
There is an optimal value $f_o$ for the given $\Delta_{qm}$ and by using the map, one can  find the optimal value $f_o$ for the given $\Delta_{qm}$.
For example  $f_o=5$ for $\Delta_{qm}=1.6$ and $\Delta_{0}=1.7$.
A variety of experimental possibilities for achieving the quantum efficiency $\eta= 20\%$ exist for a sufficiently low optical depth with narrower $\Delta_{qm}$. 
In particular, it is possible to get $\Delta_{qm}=1.2$ for $\Delta_{0}=1.3$, $f_o=3.5$ and  optical depth $\alpha_0 L\sim 8$.
Thus the optimal choice of $\Delta_{0}$ can considerably increase the spectral width $\Delta_{qm}=1.2$ of the quantum storage even for the relatively weak quantum efficiency.  

\textbf{2. Map ($\eta= 30\%$)} Fig.( \ref{Fig5}).

The map and others below is constructed similarly to the previous map. 
From the map we can find two interesting regions.
The first one relates to the atomic system with low optical depth.
Here it is reasonable  to create AFC-structure with sufficiently large $\Delta_{0}=1.2$ in order to get $\Delta_{qm}=1$ for experimentally realizable  $\alpha_0 L\geq 9$ and finesse $f=3.5$.
Wider spectral range $\Delta_{qm}=1.6$ can be also achieved but for higher optical depth $\alpha_0 L\gg 50$.  

\textbf{3. Map ($\eta= 40\%$)} Fig.(\ref{Fig6}).

Here similarly to the previous  map, we get $\Delta_{qm}=1$ for $\Delta_{0}=1.1$, $f=4$ and  $\alpha_0 L\sim 13$, .
These parameters are close to the previous case except for the higher optical depth.
Wider spectral range $\Delta_{qm}=1.5$ is possible for $\Delta_{0}=1.55$, $f=6$ and optical depth $\alpha_0 L\geq 36$.

\textbf{4. Map ($\eta= 50\%$)} Fig. (\ref{Fig7}).

Spectral range   $\Delta_{qm}=1$ for this efficiency is possible for $\Delta_{0}=1.1$, $f=4.5$a for sufficiently large optical depth  $\alpha_0 L\geq 16$. 
Thus increasing of efficiency on $10$ \% requires additional step in optical depth increase about $\delta\alpha_0 L\sim 3$ for $\Delta_{qm}=1$.   
More broadened spectral range $\Delta_{qm}=1.4$ is obtained for $\Delta_{0}=1.5$, $f=8$ and optical depth $\alpha_0 L\geq 45$.

\textbf{5. Map ($\eta= 60\%$)} Fig. (\ref{Fig8}).

Here, spectral range of quantum memory   $\Delta_{qm}=1$ is realized  for $\Delta_{0}=1.1$, $f=5$a and optical depth  $\alpha_0 L\geq 22$. 
Spectral range $\Delta_{qm}=1.3$ can be obtained  for $\Delta_{0}=1.37$, $f=7.5$ and optical depth $\alpha_0 L\geq 44$.

For brevity, we describe the properties of maps with higher quantum efficiency only by focusing on the realization of  $\Delta_{qm}=1$ and in the narrower spectral ranges due to the importance of implementation of highest quantum efficiency. 
Information on the conditions for implementing other values $\Delta_{qm}$ can be obtained from the maps given.

\textbf{6. Map ($\eta= 70\%$)} Fig.(\ref{Fig9}).
Here it is reasonable  to create AFC-structure with sufficiently large $\Delta_{0}=1.15$ in order increase up to $\Delta_{qm}=1$ for experimentally realizable  $\alpha_0 L\geq 31$ and finesse $f=5.5$.

It is interesting to note the case of $\Delta_{qm}=0.5$ for $\Delta_{0}=0.9$ (i.e. there is a quite large difference with $0.5$) $\alpha_0 L\geq 20$ and finesse $f=5.5$. 
These parameters indicate  to the realistic condition of experimental implementation  by using spectroscopic parameters of work \cite{Gundogan2013}.

\textbf{7. Map ($\eta= 80\%$)} Fig. (\ref{Fig10}).

As seen in the fig. \ref{Fig10}, spectral range $\Delta_{qm}=1$ of quantum storage is realized  for $\Delta_{0}=1.2$, finesse $f=7.5$ and optical depth $\alpha_0 L\geq 46$.
In terms of implementation of high quantum efficiency, the more interesting case is characterized by $\Delta_{qm}=0.5$, $\Delta_{0}=0.95$, finesse $f=6.5$ and optical depth $\alpha_0 L\geq 27$.
These parameters are quite close to the experimental parameters  of work \cite{Gundogan2013}.
   
\textbf{8. Map ($\eta= 90\%$)} Fig. (\ref{Fig11}).

It is worth noting that implementation of QM with so high quantum efficiency  is required for creation of quantum repeater. 
Here we note the most interesting and realistic cases of $\Delta_{qm}=0.2;0.3;0.4;0.5$ which are realized for $\Delta_{0}=1.0;1.0;1.0;1.05$ and optical depth $\alpha_0 L\geq=42;43.5;45;48$ with finesse $f=11;10;9;10$ respectively.   
Experimental implementation of these schemes requires a considerably  high optical depth that can be achieved by using a nanooptical waveguides providing a strong enhancement of the photon-atom interaction \cite{Corrielli2016, Yuan2016}.

\section{4.Conclusion} 

AFC-protocol seems to be especially interesting for implementation of broadband QM \cite{DeRiedmatten2008}.
Herein, the practical use requires a quite large quantum efficiency which still limits any application of the AFC-protocol.
It was shown \cite{Moiseev2012} that realization of highly efficient  broadband AFC-protocol should provide a strong compensation of spectral dispersion effect caused by the losses of perfect time reversibility in this type of the photon echo QM scheme.   
In the work, we have performed a detailed numerical analysis of the modified AFC-protocol (MAFC) characterized by the creation  of AFC-structure within finite spectral range of natural inhomogeneously broadened Gaussian resonant atomic transition.
Focusing to the realistic conditions for implementation of broadband AFC-protocol and its high quantum efficiency, we have constructed the maps of high spectral quantum efficiency.
The maps show a preferable condition for creation of AFC-structure providing the highest quantum efficiency within maximum broader spectral range $\Delta_{qm}$.
We characterize    the most interesting cases of modified AFC-structure and discuss its possible implementations.
The performed analysis pays the realistic ways for experimental implementation of AFC-protocol and its use for quantum repeater.

\textbf{Acknowledgements.}
The research has been supported by the the Russian Science Foundation through the Grant No. 14-12-01333-P.

\bibliographystyle{apsrev4-1}
\bibliography{bib}

\end{document}